\documentclass[10pt]{article}
\usepackage{multicol,citesort,epsfig,overcite}
\begin{document}

\title{Nucleation of a non-critical phase in a fluid near a critical point}

\author{{\bf Richard P. Sear}\\
~\\
Department of Physics, University of Surrey\\
Guildford, Surrey GU2 7XH, United Kingdom\\
email: r.sear@surrey.ac.uk}

\maketitle

\begin{abstract}
Phase diagrams of some globular proteins have a fluid-fluid
transition as well as a fluid-crystal transition.
Homogeneous nucleation of the crystal from the fluid phase
near the critical point
of the fluid-fluid transition is examined.
As the fluid-fluid critical point is approached, the number of
molecules in the critical nucleus, the nucleus at the top of the
free energy barrier to nucleation, is found to diverge as the isothermal
compressibility.
This divergence is due to a layer of the fluid phase of width
equal to the fluid's correlation length which surrounds
the core of the nucleus; the number of molecules in a crystalline
environment in the nucleus does not diverge.
The free energy barrier to nucleation remains finite but
its derivative with respect to the chemical potential is equal to minus
the number of molecules in the critical nucleus and so
diverges.
\end{abstract}

\section{Introduction}

The phase behaviour of a number of globular proteins has been studied
and in addition to the fluid-to-crystal transition there is a
metastable fluid--fluid transition
\cite{broide91,muschol97}, i.e., a transition between two
fluid phases differing in density which ends at a critical point.
This is analogous to the vapour--liquid transition of
simple substances such as water.
The transition is metastable: it exists within the fluid--crystal
coexistence region.
A fluid near the fluid-fluid critical point
is anomalous, essentially because the correlation length is very large.
Here we consider the free energy barrier $\Delta\Omega^*$
to crystallisation in a fluid which is close to the critical point.
Crystallisation starts by the nucleation
of a microscopic crystallite, this then grows to form a crystal.
The rate at which such microscopic crystallites form is proportional
to $\exp(-\Delta\Omega^*/kT)$ because their formation is
an activated process \cite{debenedetti,gunton}.
Here we study only the size and free energy
of the critical nucleus, we
leave consideration both of the dynamics of its formation and
of its growth to later work. We also limit ourselves
to temperatures above or equal to the critical temperature.
Our main finding is that the number
of molecules in the critical nucleus varies as $\chi_T$,
the isothermal compressibility, near the critical point.
As $\chi_T$ diverges at the critical point, so does the number of
molecules in the nucleus. Within our theory the number of
molecules in the critical nucleus is equal to minus the derivative
of the free energy barrier with respect to the chemical potential and
so near the critical point the barrier to nucleation is a rapidly
decreasing function of chemical potential.

We focus on the universal aspects of critical nuclei near critical
points \cite{critnote}. It is well-known that near a critical point
fluids exhibit universal behaviour, behaviour which
is determined solely by the universality
class of the system \cite{chaikin}, here that of the three-dimensional
Ising model. Although in the centre of a critical nucleus the density
will be far from the critical density,
at the fringes of the nucleus far from the centre the density
will be close to the bulk density of the fluid, and there in the
fringes we find universal behaviour. We find that the fringe of the
nucleus dominates the number of molecules in the nucleus but not
the free energy. It does however dominate the derivative of the
free energy with respect to the chemical potential
as this is nothing but minus the number of molecules in the nucleus.
In order to focus on the universal aspects we use a simple
phenomenological theory and for simplicity we use a mean-field theory,
although such theories have well-known deficiencies near critical points
\cite{chaikin}. We use a simple theory first used for droplets
by Cahn and Hilliard \cite{cahn59}. It is variously called
Cahn-Hilliard theory, van der Waals-Cahn-Hilliard theory or
the square gradient approximation
\cite{cahn59,cahn58,widom85,evans89,debenedetti}.

Numerical work on droplets near a critical point has been done
by Talanquer and Oxtoby \cite{talanquer98} using a theory similar to a
Cahn-Hilliard theory. This followed pioneering computer simulations
of a critical nucleus near a critical point by ten Wolde and
Frenkel \cite{tenwolde97}.
Talanquer and Oxtoby obtained nuclei
with very large numbers of molecules near a critical point but did
not perform the analytical analysis required to extract out the
scaling near a critical point. Although their theory
has an additional order parameter the scaling of the number of
molecules near the critical point in their model is almost
certainly the same as found here. See Refs.
\citen{haas00,dixit00,searxxx} for other recent theoretical work
on nucleation near a metastable transition.
Also, more than twenty years ago,
Widom \cite{widom85,widom77} used Cahn-Hilliard theory to look at the
(planar) interface between coexisting phases, where one of these
phases was at a critical point. He envisaged not a single
component system with fluid-fluid and fluid-crystal phase
transitions but a binary mixture with liquid-liquid demixing
and vapour-liquid phase transitions. However, the universal
aspects are the same for both systems.
Thus, our calculations for the radial
density profile of the nucleus are analogous to those for density
profile of the planar interface found by Widom. In addition,
our results are relevant to the nucleation of the vapour phase near the
critical point of a metastable liquid-liquid demixing transition.

In the next section we derive analytic expressions, within a mean-field
theory, for the density in the fringe of a nucleus in a near-critical fluid.
Then we go on to derive an expression for the number of molecules, and
hence the derivative of the free energy barrier near the critical point.
The third and final section is a conclusion.

\section{Theory}

The critical nucleus is at the top of the barrier, it is at a maximum
in the excess grand potential.
The excess grand potential is the grand potential with a nucleus minus that
without a nucleus. To start we require an expression for
the excess grand potential of a nucleus $\Delta\Omega$ as a functional
of the density function of the nucleus and then to find the density function
which extremises this functional. For a crystalline nucleus, a
rather complex functional is required, see for example the functional
used to calculate the fluid-crystal interfacial tension for hard
spheres \cite{ohnesorge94}.
However, the universal effects we are looking for here
occur at the fringe of the nucleus where the nucleus will have a density
close to that of the fluid. There a much simpler functional suffices,
such as those
used to calculate fluid-fluid interfacial tensions and the excess
grand potentials of fluid droplets in fluid phases.

For a spherically symmetric fluid droplet in a fluid, the
density profile $\phi(r)$ is a function only of the distance $r$
from the centre of the droplet. The fluid
phase has a density $\rho$ and a chemical potential $\mu$;
$\phi$ is the density at a point
whereas $\rho$ is the bulk density of the fluid. Of
course $\phi(r\rightarrow\infty)=\rho$.
The standard Cahn-Hilliard expression for the
grand potential cost $\Delta\Omega$ of such a droplet, as a functional
of its density profile $\phi(r)$, is
\cite{cahn58,cahn59,widom85,evans89,debenedetti}
\begin{equation}
\Delta\Omega = \int\left[
\Delta\omega + \kappa\left(\nabla\phi\right)^2\right]
{\rm d}{\bf r},
\label{cahn}
\end{equation}
where
\begin{equation}
\Delta\omega(\phi)=f(\phi)-f(\rho)-\mu(\phi-\rho),
\label{omega}
\end{equation}
is the work required per unit volume to change the density from $\rho$
to $\phi$, at a chemical potential $\mu$.
$f(\phi)$ is the bulk Helmholtz free energy per unit volume of the
fluid at a density $\phi$.
The second term in Eq. (\ref{cahn})
within the brackets is the gradient term: the
grand potential cost due to variations in space of the density.
The gradient squared term is the lowest order term in a gradient
expansion and so is adequate when the density is slowly
varying. The coefficient, $\kappa$ of this term is
taken to be a constant. See
Refs. \citen{widom85,evans89,debenedetti} for its relation to the
intermolecular potential.

The functional Eq. (\ref{cahn}) will be totally inadequate within the
crystalline core of a nucleus but we will not require either the density
function or the contribution to the grand potential of this core.
Thus, we will use Eq. (\ref{cahn}) as our functional and end up
with expressions which are the sum of two terms, one term from the core
for which our expression will be wrong but which we will not
evaluate, and another term from the fringe, for which Eq. (\ref{cahn})
is a reasonable (mean-field) approximation which we will evaluate.

The critical nucleus is at the top of the free energy barrier
and so is at a maximum of $\Delta\Omega$. Thus, for the
critical nucleus we may set the functional derivative of
$\Delta\Omega$ with respect to the density profile $\phi(r)$
to zero,
\begin{equation}
\left(\frac{\partial \Delta\omega}{\partial\phi}\right)-2\kappa\nabla^2\phi
=0.
\label{temp}
\end{equation}
The curvature of the density profile at a point is proportional
to the derivative of the excess grand potential with respect to
the density at that point. Using Eq. (\ref{omega}) for $\Delta\omega$
in Eq. (\ref{temp}) we have
\begin{equation}
\mu(\phi)-\mu-2\kappa\nabla^2\phi =0,
\label{max}
\end{equation}
where the $\mu$ without an argument is the
chemical potential in the fluid, and $\mu(\phi)$ is the chemical
potential of a bulk fluid at a density $\phi$.
Once we have solved Eq. (\ref{max}) we can insert the solution
into Eq. (\ref{cahn}) to obtain the excess grand potential of the
critical nucleus, denoted by $\Delta\Omega^*$.

The excess number of molecules in the critical nucleus, $\Delta n^*$,
is the number of molecules with the nucleus present minus the number
without it. It may be obtained by integrating over the density profile,
\begin{equation}
\Delta n^*=
\int \Delta\phi(r) {\rm d}{\bf r},
\label{nstardef}
\end{equation}
where $\Delta\phi(r)=\phi(r)-\rho$ is the excess density at a point,
the density at a point minus the bulk density.
The derivative of $\Delta\Omega^*$ with respect
to the chemical potential of the fluid $\mu$ is, using
Eq. (\ref{cahn}),
\begin{equation}
\left(\frac{\partial\Delta\Omega^*}{\partial\mu}\right)_T=
-\int \Delta\phi(r) {\rm d}{\bf r}
=-\Delta n^*.
\label{nstar}
\end{equation}
Thus, the derivative of $\Delta\Omega^*$ with respect to
the chemical potential is simply $-\Delta n^*$.
This result, that the
derivative of the excess grand potential of the critical nucleus
is minus the excess number of molecules in the nucleus,
is often called the nucleation theorem
\cite{kashchiev82,viisanen93,bowles00}.

\subsection{The fringe of the nucleus}

\begin{figure}
\begin{center}
\caption{
\lineskip 2pt
\lineskiplimit 2pt
A schematic of the radial density profile of a droplet:
$\phi(r)$ is the density at a distance $r$ from the centre of the
droplet. The density of the surrounding fluid is $\rho$, and
$r_c$ is the (somewhat arbitrary) point where we divide
the density profile into a central part and an outer part.
}
\vspace*{0.1in}
\epsfig{file=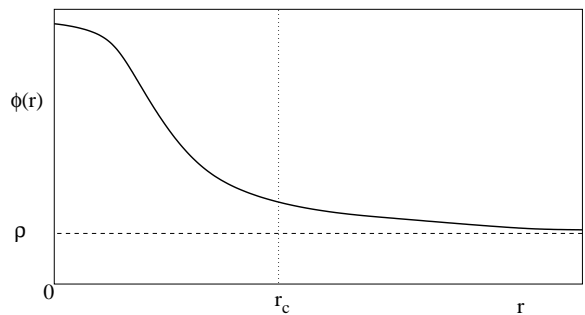,width=3.0in}
\end{center}
\label{figprof}
\end{figure}

Equation (\ref{max}) can be solved numerically if the chemical potential
is known as a function of density. Here, we would like to
concentrate on the fringe of the critical nucleus. The
fringe is the outermost part of the nucleus, where 
the density is near the density of the fluid, $\rho$.
We define the fringe as being that part of the nucleus which
is more than a distance $r_c$ from the centre.
The distance $r_c$ is such that $\Delta\phi(r\ge r_c)/\rho\ll 1$.
Fig. \ref{figprof} is
a schematic of the radial density profile of the nucleus.
Now, if the fractional density difference $\Delta\phi/\rho\ll 1$
we can use a Taylor expansion for $\mu(\phi)-\mu$,
\begin{equation}
\mu(\phi)-\mu=\Delta\phi
\left(\frac{\partial\mu(\phi)}{\partial\phi}\right)_{\phi=\rho}+\cdots
=\frac{\Delta\phi}{\rho^2\chi_T}+\cdots,
\label{taylor}
\end{equation}
where $\chi_T$ is the isothermal compressibility of the fluid (at
a density $\rho$). The isothermal compressibility is defined as
\cite{hansen86}
\begin{equation}
\chi_T^{-1}=\rho^2\left(\frac{\partial\mu}{\partial\rho}\right)_{V,T}.
\label{chi}
\end{equation}
Substituting Eq. (\ref{taylor}) into Eq. (\ref{max}) we have
\begin{equation}
\frac{1}{\rho^2\chi_T}\Delta\phi(r)-
2\kappa\nabla^2\Delta\phi(r) =0 ~~~~~~~ \Delta\phi\ll\rho.
\label{helm}
\end{equation}
This is the Helmholtz equation, for $\Delta\phi$,
and the solution is a function
of the Ornstein-Zernike form,
\begin{equation}
\Delta\phi(r)=\rho\frac{\sigma}{r}\exp(-r/\xi),
\label{oz}
\end{equation}
with $\xi$ the correlation length of the fluid, given by
\begin{equation}
\xi^2=2\kappa\rho^2\chi_T.
\label{xi}
\end{equation}
To obtain Eq. (\ref{oz}) the boundary conditions
$\Delta\phi(r\rightarrow\infty)\rightarrow0$ and
$\Delta\phi(r_c)=\rho(\sigma/r_c)\exp(-r_c/\xi)$
were employed. $\sigma$ is a molecular length scale, a few nms for
proteins. $\Delta\phi/\rho$ will, as required
for Eq. (\ref{helm}), be small for $r\ge r_c$ provided
that $r_c$ is a few times $\sigma$ or more. We do not need to specify
$r_c$ beyond saying that it must be at least a few times $\sigma$.
This implies a core a few molecules across, which is reasonable.
From Eq. (\ref{oz}) we see that
the width of the fringe is, as we might have expected,
of the order of the correlation length $\xi$ in the surrounding fluid.
This width will thus diverge as the fluid approaches a critical point.

\subsection{Near a critical point}

We now consider a nucleus in a fluid which is near a critical point,
either at the critical density but just above the critical temperature,
or at the critical temperature but at a density near the
critical density. In either case as the critical point is approached the
isothermal compressibility, and
hence (see Eq. (\ref{xi})) $\xi$, diverges. Within mean-field theory,
along the critical isochore, i.e., with the density fixed at its value
at the critical point $\rho_{cp}$, $\chi_T$ scales with temperature
as \cite{chaikin,debenedetti}
\begin{equation}
\chi_T \sim (T-T_{cp})^{-1}  ~~~~~~~~~\rho=\rho_{cp}  ~~~T>T_{cp},
\label{isoch}
\end{equation}
where $T_{cp}$ is the temperature of the critical point.
The compressibility diverges as one over the
temperature difference to the critical point. An alternative
path to the critical point is along the critical isotherm.
We fix the temperature $T=T_{cp}$ and vary the density. Along the
critical isotherm $\chi_T$ varies as
\begin{equation}
\chi_T \sim (\rho-\rho_{cp})^{-2}  ~~~~~~~~~ T=T_{cp}.
\label{isoth}
\end{equation}
The compressibility diverges as one over the square of the
density difference to the critical point.

Putting our solution for $\Delta\phi(r>r_c)$, Eq. (\ref{oz}),
into Eq. (\ref{nstardef})
we find that when $\xi$ is very large and so
the fringe correspondingly large in volume, that the fringe
dominates $\Delta n^*$. Equation
(\ref{nstardef}) is easily evaluated by substituting Eq.
(\ref{oz}) for $\Delta\phi(r>r_c)$,
\begin{eqnarray}
\Delta n^*
&=&
\int_{r=0}^{r=r_c}
\Delta\phi(r) {\rm d}{\bf r}
+4\pi\rho\sigma\xi^2~~~~~\xi\gg r_c
\nonumber\\
&=&
\int_{r=0}^{r=r_c}
\Delta\phi(r) {\rm d}{\bf r}
+8\pi\sigma\kappa\rho^3\chi_T~~~~~\xi\gg r_c,
\label{ncrit}
\end{eqnarray}
where we used the relation between $\xi$ and $\chi_T$, Eq. (\ref{xi}).
The excess number of molecules $\Delta n^*$ is a sum of two terms.
The first term, the integral in Eq. (\ref{ncrit}), comes from
the core of the nucleus. It is of order
$(\rho_c-\rho)r_c^3$ where $\rho_c$ is the density of the
phase which is nucleating. It remains finite as the critical
point is approached. However, the second term
in Eq. (\ref{ncrit}), the contribution
of the fringe to the number
of molecules in the critical nucleus $\Delta n^*$, diverges.
As the critical point is approached the number of molecules
in the nucleus tends to infinity, as suggested by
Talanquer and Oxtoby \cite{talanquer98}. It does so as the compressibility
$\chi_T$. This is the central result of this work, and
is a general result independent of the nature of the phase which is
nucleating.
The fringe of the nucleus is at densities close to the fluid's
density and far from that in the core of the nucleus. So,
the divergence in $\Delta n^*$ is from a divergent number of molecules
at densities near the bulk fluid density: although they are part of
a nucleus of a crystalline phase they themselves are in a fluid
environment.
The number of molecules which are in a crystalline environment
does not diverge as the critical point is approached.
Thus, within mean-field theory, along the critical
isochore the number of molecules diverges as
$(T-T_{cp})^{-1}$ and along the critical isotherm as ($\rho-\rho_{cp})^{-2}$.

Now we consider the contribution of the fringe to the grand potential
of the critical nucleus. We require the Taylor
expansion of $\Delta\omega$ about its value at the bulk density,
\begin{equation}
\Delta\omega(\phi)=\frac{1}{2}
\frac{(\Delta\phi)^2}{\rho^2\chi_T}+\cdots.
\label{quad}
\end{equation}
The first nonzero term in the Taylor expansion of $\Delta\omega$
is the quadratic term as the first density derivative of $\Delta\omega$
is zero at $\rho$. Now, splitting the integration in Eq. (\ref{cahn})
at $r_c$, and using the quadratic approximation, Eq. (\ref{quad}),
for $\Delta\omega$ for $r>r_c$, we have
\begin{equation}
\Delta\Omega =
\int_{r=0}^{r=r_c}\left[
\Delta\omega + \kappa\left(\nabla\phi\right)^2\right]{\rm d}{\bf r}+
\int_{r=r_c}^{r=\infty}\left[\frac{1}{2}
\frac{(\Delta\phi)^2}{\rho^2\chi_T}
+ \kappa\left(\nabla\Delta\phi\right)^2\right]
{\rm d}{\bf r},
\label{cahnlin}
\end{equation}
where, for $r>r_c$,
we also substituted $\Delta\phi$ for $\phi$ in the gradient term.
After substituting Eq. (\ref{oz}) for $\Delta\phi$ in the
second integral,
we see that the contributions to the integral from both the
terms in the integrand {\em decreases} as a function of radial
distance $r$. Taking into account the factor of $r^2$
in ${\rm d}{\bf r}$ the integrand varies as $\exp(-2r/\xi)$ times a quadratic
polynomial in $1/r$: it is a monotonically decreasing function of $r$.
The farther we go out into the fringe of the nucleus the
smaller is the contribution to the excess grand potential,
the barrier to nucleation. This means that the barrier to nucleation
will be dominated by contributions from the core of the nucleus
where the behaviour is non-universal and our approximations are invalid.
Recall that our functional, Eq. (\ref{cahn}), is a very
poor approximation in the core and hence the first
integral in Eq. (\ref{cahnlin}) is not an accurate expression for
the contribution of the core to the excess grand potential.
We are thus unable to calculate the absolute excess
grand potential of a critical nucleus
near a critical point but we are of course able to calculate its
derivative with respect to the chemical potential as this is just
minus the excess number of molecules in the critical nucleus,
which is dominated by the fringe.

\section{Conclusion}

Some globular proteins \cite{broide91,muschol97} have two transitions:
a fluid-crystal transition, call it transition $\alpha$, and
a fluid-fluid transition, call it transition $\beta$. Transition
$\alpha$ is always strongly first order but transition $\beta$
becomes continuous at the critical point. Here, we have studied
homogeneous nucleation associated with transition $\alpha$ near
a continuous transition $\beta$.
Systems near a continuous transition
show universal (within a universality class)
behaviour and are very susceptible to a perturbation --- 
response functions such as the compressibility take large values.
The nucleus associated with homogeneous nucleation of a
transition $\alpha$ therefore perturbs the highly
susceptible fluid near transition $\beta$.
Because the fluid is near this continuous
transition, the response functions such as the isothermal
compressibility $\chi_T$, are large. The perturbation due to the
nucleus is therefore large, it extends out to the large distance $\xi$ and
involves a number of molecules which scales with $\chi_T$.
The free energy barrier to homogeneous nucleation of transition
$\alpha$ is then a rapidly decreasing function of chemical potential.
As suggested by ten Wolde and Frenkel \cite{tenwolde97}, the
continuous transition $\beta$ helps the dynamics of transition
$\alpha$.
This is true in general, not just for crystallisation near
a critical point. The nucleation of the vapour phase when 
a mixture of liquids boils near a metastable critical point of liquid-liquid
demixing is another example. Note that there $\Delta n^*$ is negative
and will tend towards $-\infty$ not $+\infty$.

The theory of section 2 is a simple mean-field theory. However,
our conclusion that $\Delta n^*\sim\chi_T$ near the critical
point may be valid beyond mean-field theory. If so then
the divergence of $\Delta n^*$ along a particular path in the
diagram, e.g., the critical isochore, will occur with whatever is the 
correct critical exponent of $\chi_T$ along that path.
For example, for nucleation near a
critical point in the universality class of the Ising model
in three dimensions, $\chi_T$ scales as
$(T-T_{cp})^{-1.24}$ along the critical isochore not
$(T-T_{cp})^{-1}$ as mean-field theory predicts, see
Refs. \citen{debenedetti,chaikin}.

The starting point for most theoretical studies of nucleation is
classical nucleation theory \cite{debenedetti,gunton}.
Within classical
nucleation theory the  critical nucleus is modeled as a small crystallite,
assumed to have bulk properties, separated from the surrounding
metastable fluid by an interface, assumed to have the same surface
tension as that between the coexisting bulk phases and to be thin,
no more than a couple of molecules thick. For a nucleus in a fluid
far from a fluid-fluid critical point, this assumption of a thin
interface with the bulk surface tension is not unreasonable.
The correlation lengths in the fluid and crystalline phases
will both be small and so will the width of the interface. Then if
the width of the interface is small with respect to the diameter
of the nucleus then approximating its surface tension by that
of the planar interface between coexisting phases is reasonable.
However, for a nucleus in a near-critical fluid, the correlation
length in the surrounding fluid $\xi$ may be very large.
So, for a crystal nucleus which is only a few lattice spacings
across the interface between it and the surrounding
fluid may be much thicker than the diameter of the crystalline
part of the nucleus. Then this interface will be very different from
that assumed within classical nucleation theory, and
so classical nucleation theory will yield only a poor estimate
of the barrier to nucleation.

It is a pleasure to thank R. Bowles for
useful discussions.
Work supported by EPSRC (GR/N36981).


\begin{thebibliography}{99}

\bibitem{broide91} M. L. Broide, C. R. Berland, J. Pande, O. O. Ogun
and G. B. Benedek, Proc. Nat. Acad. Sci. {\bf 88}, 5660 (1991).

\bibitem{muschol97} M. Muschol and F. Rosenberger,
J. Chem. Phys. {\bf 107}, 1953 (1997).

\bibitem{debenedetti} P. G. Debenedetti,
{\it Metastable Liquids}
(Princeton University Press, Princeton, 1996).

\bibitem{gunton} J. D. Gunton, M. San Miguel and P. S. Sahni,
in {\it Phase Transitions} volume 8, edited by C. Domb and J. L. Lebowitz
(Academic Press, London, 1983).

\bibitem{critnote}
Note that conventionally `critical'
is used to denote both a point in the phase diagram and the crystallite
at the top of the barrier although there is no connection between the
two uses of the word. Although this terminology is a little
unfortunate we will use it here and so we will be studying
a critical nucleus near a critical point.

\bibitem{chaikin} P. M. Chaikin and T. C. Lubensky,
{\it Principles of Condensed Matter Physics}
(Cambridge University Press, Cambridge, 1995).

\bibitem{cahn59} J. W. Cahn and J. E. Hilliard,
J. Chem. Phys. {\bf 31}, 688 (1959).

\bibitem{cahn58} J. W. Cahn and J. E. Hilliard,
J. Chem. Phys. {\bf 28}, 258 (1958).

\bibitem{widom85} B. Widom,
Chem. Soc. Rev. {\bf 14}, 121 (1985).

\bibitem{evans89} R. Evans,
{\it Les Houches, Session XLVIII: Liquids at Interfaces},
edited by J. Charvolin, J. F. Joanny and J. Zinn-Justin
(Elsevier, Amsterdam, 1989)

\bibitem{talanquer98} V. Talanquer and D. W. Oxtoby,
J. Chem. Phys. {\bf 109}, 223 (1998).

\bibitem{tenwolde97} P. R. ten Wolde and D. Frenkel,
Science {\bf 277}, 1975 (1997).

\bibitem{haas00} C. Haas and J. Drenth,
J. Phys. Chem. B {\bf 104}, 368 (2000).

\bibitem{dixit00} N. M. Dixit and C. F. Zukoski,
J. Coll. Int. Sci. {\bf 228}, 359 (2000).

\bibitem{searxxx} R. P. Sear,
cond-mat/9912199 (http://xxx.lanl.gov).

\bibitem{widom77} B. Widom,
J. Chem. Phys. {\bf 67}, 872 (1977).

\bibitem{ohnesorge94} R. Ohnesorge, H. L\"{o}wen and H. Wagner,
Phys. Rev. E {\bf 50}, 4801 (1994).

\bibitem{kashchiev82} D. Kashchiev,
J. Chem. Phys. {\bf 76}, 5098 (1982).

\bibitem{viisanen93} Y. Viisanen, R. Strey and H. Reiss,
J. Chem. Phys. {\bf 99}, 4680 (1993).

\bibitem{bowles00} R. K. Bowles, R. McGraw, P. Schaaf, B. Senger,
J.-C. Voegel and H. Reiss,
J. Chem. Phys. {\bf 113}, 4524 (2000).

\bibitem{hansen86} J.-P. Hansen, and I. R. McDonald,
{\it Theory of Simple Liquids} (Academic Press, London, 2nd Edition, 1986).


\end{thebibliography}
\end{document}